\documentclass[onecolumn,showpacs,showkeys,preprintnumbers,amsmath,amssymb,superscriptaddress,prstab]{revtex4}
\usepackage{graphicx}% Include figure files
\usepackage{dcolumn}% Align table columns on decimal point
\usepackage{bm}% bold math
\usepackage{xcolor}
\usepackage{mathtools}
\usepackage{algorithm}
\usepackage{algpseudocode}
\usepackage{amsmath, amsthm, amssymb}
\usepackage{subfig}
\usepackage{lscape}
\usepackage{booktabs}% http://ctan.org/pkg/booktabs
\newcommand{\mb}[1]{\mathbf{#1}}

\renewcommand{\vec}[1]{{\boldsymbol{#1}}}

\begin{document}
\title{Accurate numerical  simulation of radiation reaction effects \\
in strong  electromagnetic fields} 
\author{N.V.~Elkina}%
\affiliation{Ludwig-Maximilians Universit\"{a}t, M\"unchen, 80539, Germany}
\email{Nina.Elkina@physik.uni-muenchen.de}
\author{A.M.~Fedotov}
\affiliation{National Research Nuclear University MEPhI, Moscow, 115409, Russia}
\author{C.~Herzing}
\affiliation{Ludwig-Maximilians Universit\"{a}t, M\"unchen, 80539, Germany}
\author{H.~Ruhl}
\affiliation{Ludwig-Maximilians Universit\"{a}t, M\"unchen, 80539, Germany}

\begin{abstract} 
The Landau-Lifshitz equation provides an efficient way  to account for the effects of radiation reaction without acquiring the non-physical solutions typical for the Lorentz-Abraham-Dirac equation. We solve the Landau-Lifshitz equation in its covariant four-vector form  in order to control both the energy and momentum of radiating particle. Our study reveals that implicit time-symmetric collocation methods of the  Runge-Kutta-Nystr\"om type are superior in both accuracy and better maintaining the mass-shell condition than their explicit counterparts. We carry out an extensive  study of numerical accuracy by comparing the  analytical and numerical  solutions of the Landau-Lifshitz equation. Finally, we present the results of simulation of  particles scattering by a  focused laser pulse. Due to radiation reaction, particles are less capable for penetration into the focal region, as compared to the case of radiation reaction neglected. Our  results are important for designing the forthcoming experiments  with high intensity laser fields. 
\end{abstract}
\maketitle

\section{Introduction}
In recent years much interest has been attracted to the problem of radiation reaction in a realm of high intensity laser fields. Rigorous approach to radiation reaction requires  quantum treatment of the stochastic events of hard photon emission \cite{Nikishov:1967, Baier:1969}. In numerical simulation this was incorporated by using Monte-Carlo sampling of photon emission \cite{Elkina:2011}. This  method treats correctly the recoil due to emission of hard photons with $\hbar\omega > mc^2\sim 0.511$ MeV ($m$ is the electron mass, and $c$ is the speed of light) but is computationally rather expensive. At the same time, the relatively soft part of the photon spectrum $10$keV - $0.511$MeV had to be ignored in plasma simulations due to a limited numerical resolution. 

One of the approaches to account for radiation effect due to soft photon emission is to consider it classically by adding the radiation friction term into the  equation of particle motion. The classical problem of motion of a radiating charge was considered in \cite{Dirac:1938} where the so-called Lorentz-Abraham-Dirac (LAD) equation was derived. This equation of motion  contains third order derivative $\dddot{x}^\mu$ and therefore possesses problematic run-away or causality violating solutions. To ensure physical behaviour the LAD equation requires careful choice of both initial and boundary conditions (see \cite{Klepikov:1985}, \cite{Rohrlich:2007} for review) which is difficult to realize  in computational practice. In the most of physically meaningful situations radiation reaction can be treated as a perturbation. The first order of perturbation theory results in the Landau-Lifshitz (LL) equation \cite{Landau:II}. This equation contains nonlinear dissipation  which leads to contraction of a phase space volume. Proper numerical method should capture this effect. 

In plasma simulation solvers \cite{Birdsal:2004} the equations of particle motion  are usually integrated by the modified leap-frog (St\"ormer-Verlet) method. This originally semi-implicit method can be turned to an explicit one provided that the special ansatz \cite{Boris:1970, Vay:2008} is used to treat properly the rotation in a magnetic field. However, inclusion of an  extra term responsible for the radiation reaction breaks the symmetry of the relativistic Newton-Lorentz (NL) equation. Since radiation reaction depends upon momenta nonlinearly, the explicitness of the method can be no more maintained within the second order of accuracy. To avoid the loss of accuracy, one may consider  different higher order methods,  primarily the single-step Runge-Kutta (RK) like  methods \cite{Butcher:2008}. 
  
Another concern with simulation of the covariant LL equation is  maintaining the mass shell condition (conservation of the Minkowski norm $u_\mu u^\mu = c^2$). The Minkowski norm can be preserved exactly in a course of numerical simulation if a corresponding equation of motion is lifted onto the Lie group settings \cite{Kawaguchi:2003, Harvey:2011}. The Lie group methods show promising results, however are rather expensive for nonlinear equations, as they require multiple computations of the matrix exponentials for each time step \cite{Moler:1978}.   

In this paper we would like to draw attention to the Runge-Kutta-Nystr\"om [RK(N)] type methods \cite{Nystrom:1925}, which are specially designed for solving the second order ordinary differential equations (ODEs). However, the explicit RK type methods do not conserve exactly the quadratic integrals of motion (e.g., energy and the Minkowski norm) and hence are not enough suitable for integration over a long time. Contrarily, the implicit collocation RK(N) methods \cite{Hairer:2002} provide better accuracy and are capable for conservation of quadratic integrals of motion. This, together with time reversibility, singles out these methods as perfect candidates for integration of the LL equation over a long time. 

The paper is organized as follows. In Section~\ref{basic} a summary of covariant formulation of the equation of motion of radiating particle is presented. The main  goal  of the paper is achieved in Section~\ref{numeric}, where we present derivation of the collocation Gauss-Legendre RKN method entirely in terms of collocation points for fourth, sixth and eighth orders of accuracy. The result is employed in Sections~\ref{mag_test} and~\ref{pw_test}, where we perform extensive validation studies of the considered RK(N) methods against exact solutions of the LL equation in a constant magnetic field and in a plane wave and demonstrate that implicit collocation methods of RK(N) type are superior in accuracy to their explicit counterparts in solution of both the NL and LL equations. In Section~\ref{pond} we present an important application of the developed numerical technique to the problem of relativistic scattering of electrons by a  focused laser field. Our conclusions are summarized in Section~\ref{conclusion}.

\section{Basic equations}
\label{basic}
Radiation reaction modifies the NL equation by a radiation friction term $g^\mu$ in the right hand side 
\begin{align}
\frac{d^2 x^\mu}{d\tau^2} = f^\mu= \frac{e}{m} F^\mu_\nu u^\nu + g^\mu.
\label{motion}
\end{align}
The 4-velocity $u^\mu = \{\gamma,\; \mb u\}$ of a charge is the derivative of the position 4-vector $\mb x^\mu=\{t,\; \mb r\}$ with respect to the proper time $\tau$. 
The electromagnetic field tensor $F^\mu_\nu$ is expressed by the following matrix 
\begin{align}
F^\mu_\nu = \left[
\begin{array}{cccc}
0 &E_x &E_y &E_z\\
E_x &0 &B_z &-B_y\\
E_y &-B_z&0 &B_x\\
E_z &B_y&-B_x &0
\end{array}
\right]\nonumber
\end{align}
Radiation reaction was originally considered with the LAD equation \cite{Dirac:1938}
\begin{align}
\frac{d^2 x^\mu}{d\tau^2} = \frac{e}{m} F^\mu_\nu u^\nu + \frac{2}{3}\frac{e^2}{m} \left(u_\nu\frac{d^2 u^\mu}{d\tau^2}- u^\mu\frac{d^2 u_\nu}{d\tau^2}\right) u^\nu,
\label{LAD}
\end{align}
here and in the following we adopt the units $c=\hbar=1$, $e=\sqrt{\alpha}$, $\alpha=e^2/\hbar c \simeq 1/137$ is the fine structure constant. 
As was pointed out in the Introduction, the equation~(\ref{LAD}) possesses the problematic run-away or causality violating solutions. The regular way to avoid such unwanted solution is to treat the second term in right-hand-side as a perturbation \cite{Landau:II}. Then in virtue of 
\begin{align}
\dot{F}_\nu^\mu=\frac{d x^\lambda}{d\tau}\frac{\partial F_\nu^\mu}{\partial x^\lambda}=u^\lambda\partial_\lambda F^\mu_\nu \nonumber
\end{align}
the term $g^\mu$ takes the form 
\begin{align}
g^\mu = \frac{2e^3}{3m^2}u^\lambda (\partial_\lambda F^{\mu}_\nu) u^\nu  + \frac{2e^4}{3m^3}\left[ F^{\mu}_{\nu}F^{\nu}_\lambda u^\lambda + (F_{kl}u^l)(F^k_mu^m) u^\mu\right].
\label{gl}
\end{align} 
Eq.~(\ref{motion}) with the term $g^\mu$ given by~(\ref{gl}) is known as the LL equation. The numerical solution of the LL equation should preserve the Minkowski norm $u^\mu u_\mu = 1$, as well as the orthogonality condition $g^\mu u_\mu = 0$. It turns out that actual fulfilment of the former condition strongly depends on a chosen numerical method, while implementation of the latter one is difficult to enforce if $g^\mu$ is in the form~(\ref{gl}). Indeed, in this form $g^\mu$ consists of three terms, which contribute disparately. While in the majority of situations the first term in~(\ref{gl}) can be safely neglected, the two remaining terms together do satisfy the  orthogonality condition, but due to a huge difference in the magnitude of their components naive summation would lead to  accumulation of considerable round-off errors. This purely numerical problem may cause eventual violation of orthogonality between the 4-force and 4-momentum. In order to avoid such an unwanted effect we consider the LL equation in the ``symmetrized'' form reminiscent of the LAD equation. So that the actual equation which is solved in this paper is given as follows
\begin{align}
\frac{d^2 x^\mu}{d\tau^2 }  &= \frac{e}{m} F^\mu_\nu u^\nu + 
\frac{2}{3}\frac{e^2}{m}\left(u_\nu\dot{ w}^\mu-u^\mu\dot{w}_\nu\right) u^\nu,
\label{sym}
\end{align}
where  $\dot{w}^\mu$ is calculated as
\[
\dot{w}^\mu = \frac{e}{m}\left(
\dot{F}^\mu_\nu u^\nu+F^\mu_\nu\dot{u}^\nu\right)=\frac{e}{m}\left[u^\lambda(\partial_\lambda F^\mu_\nu) u^\nu+\frac{e}{m}F^\mu_\nu F^\nu_\lambda u^\lambda\right].
\] 
In Eq.~(\ref{sym}) both leading radiation reaction terms acquire the same order,
so that their subtraction can no longer cause accumulation of round-off errors. Such a trick allows to keep $g^\mu u_\mu=0$ satisfied up to the level of machine precision.
%%%%%
\section{Numerical integration of  the motion of  a charge}
\label{numeric}
\subsection{Explicit Runge-Kutta-Nystr\"om methods}
The most frequently preferred approach to solve a second order ODE $\ddot{x}^\mu=f^\mu(\tau,x^\mu,\dot{x}^\mu)$ with higher accuracy is to apply a suitable method to an equivalent system of the first order system of ODEs
\begin{align}
\frac{d}{d\tau}\left(
\begin{array}{c}
x^\mu\\
u^\mu
\end{array}
\right) = \left(
\begin{array}{c}
u^\mu\\
f^\mu(\tau,\; x^\mu,\; u^\mu)
\end{array}
\right).
\end{align}
Application of a RK method to such a problem  results in \footnote{The time step index $n$ of the auxiliary stage quantities $K_i^\mu$, $L_i^\mu$ is assumed but omitted for simplicity sake.} 
\begin{align}
L^\mu_i &= u_n^\mu +h\sum_{j=1}^s A_{ij}K^\mu_j,\quad  i=1,...,s, \label{erk1_1}\\
K^\mu_i &= f^\mu\left(\tau_n + h c_i,\;  x^\mu_n + h\sum_{j=0}^s A_{ij} L_{j},\;  
u^\mu_n + h\sum_{j=1}^s A_{ij}K_j\right),\quad  i=1,...,s,\label{erk1_2}\\
x^\mu_{n+1} &= x^\mu_n + h\sum_{j=1}^s a_j L_j^\mu,\label{erk1_3}\\
u^\mu_{n+1} & = u^\mu_n + h\sum_{j=1}^s a_j K_j^\mu,\label{erk1_4}
\end{align}
where $h$ is a step size and  $A_{ij}$, $c_i$ and $a_i$ are the scheme-specific coefficients (see Table~\ref{trk}). 
\begin{table*}[!ht]
\caption{The Butcher tables for RK and RKN methods.}\label{tt}\addtocounter{table}{-1}
%\centring
\subfloat[RK]{\label{trk}
  \begin{tabular}[b]{c|c} %\hline
   $\vec c$ &$A_{s\times s}$ \\\hline
   	     &$\vec a^T$
   \end{tabular}
   }
   \hspace{4em}
   %%%%%%
\subfloat[RKN]{\label{trkn}
       \begin{tabular}[b]{c|c|c} %\hline
   $\vec c$ &$\mb A_{s\times s}$ &$\mb B_{s\times s}$ \\\hline
   	     &$\vec a^T$ & $\vec b^T$
   \end{tabular}
   }\addtocounter{table}{1}
  \end{table*} 
An alternative approach to solve the equation of motion is to explore the original second order ODE $\ddot{x}^\mu=f^\mu(\tau,x^\mu,\dot{x}^\mu)$ without prior reducing it to a system of the first order ODEs. This can be achieved by eliminating $L_i$ from  the formulas~(\ref{erk1_1}) - (\ref{erk1_4}) as follows 
\begin{align}
 K_i^\mu &= f^\mu\left(\tau_n+c_ih, x^\mu_n + c_ih u^\mu_n + h^2\sum_{j=1}^s B_{ij}K_j^\mu, u^\mu_n + \sum_{j=1}^s A_{ij}K_j^\mu\right),\label{erk2_1}\\
 x^\mu_{n+1} &= x_n^\mu + h u^\mu_n + h^2\sum_{j=1}^s b_j K_j^\mu,\label{erk2_2}\\
 u^\mu_{n+1} & = u_n^\mu + h\sum_{j=1}^s a_jK_j^\mu,\label{erk2_3}
 \end{align}
where the coefficients $B_{ij}$ and $b_{j}$ are given by 
 \begin{align}
 B_{ij} = \sum_k A_{ik}A_{kj},\quad b_j = \sum_k a_kA_{kj}.
 \label{cond_rkn}
 \end{align}
The conditions~(\ref{cond_rkn}) for the new coefficients $B_{ij},\;b_j$ in fact can be abandoned, as it was found by Nystr\"om in \cite{Nystrom:1925}. Such kind of methods is known as the RKN methods, see also \cite{Fehlberg:1975}. In most general form the RKN methods are defined by Eqs.~(\ref{erk2_1}) - (\ref{erk2_3}), where the coefficients $A_{ij}$, $B_{ij}$, $a_i,\; b_i$ are kept independent and form the double Butcher array (see Table~(\ref{trkn})).
\begin{table*}[!ht]
  \caption{The explicit RK(N) methods of order $4$.}
\subfloat[RK]{
\begin{tabular}{c|cccc}
0&0&0&0&0\\
1/2&1/2&0&0&0\\
1/2&0&1/2&0&0\\
1             &0&0&1&0\\\hline
              &1/6&1/3&1/3&1/6
\end{tabular}
   }
   \hspace{4em}
   %%%%%%
\subfloat[RKN]{
\begin{tabular}{c|cccc|cccc}
0    &0&0&0&0                      &0&0&0&0\\
1/2&1/2&0&0&0        &1/8&0&0&0\\
1/2&0&1/2&0&0  &1/8&0&0&0\\
1              &0&0&1&0   &0&0&1/2&0\\\hline
&1/6&1/3&1/3&1/6			&1/6&1/6&1/6&0   
 \end{tabular}
   }
   \label{erk}
  \end{table*} 
If  $A_{ij} =0$ and $B_{ij} = 0$  for $j\geq i$, then such construction yields an explicit method, in which each stage $K_i^\mu$ can be determined sequentially. The examples of explicit forth order RK and RKN methods are given in Table~\ref{erk}. 

\subsection{Implicit collocation Runge-Kutta-Nystr\"om methods} 
\label{irk}
The basic idea of the collocation methods  \cite{Hairer:2002} is to select such  polynomials $r^\mu(\tau)$ of degree $s+1$ that satisfy the initial value problem at $s+1$ points $\tau_n,\,\tau_n+c_1 h,\,\ldots,\,\tau_n+c_s h\in [\tau_n,\tau_{n+1})$, where $c_1,\ldots,c_s$ are the distinct real numbers from the interval $(0,\;1)$. 
Then for the second order ODE $\ddot{x}^\mu=f^\mu(\tau,x^\mu,\dot{x}^\mu)$ the 
corresponding polynomial approximation $r^\mu$ satisfies \cite{Brunner:1987}
\begin{align}
r^\mu(\tau_n) =x^\mu_n,\quad
\dot r^\mu(\tau_n) =u^\mu_n,\quad
\ddot r^\mu(\tau_n + c_i h) = f^\mu(\tau_n + c_i h,\; r^\mu(\tau_n + c_i h),\;\dot r^\mu(\tau_n + c_i h))\quad [i=1,...,s].\label{rkn11}
\end{align}
In the following we proceed with a derivation of an RKN method (the corresponding RK method is  defined by a subtable with the coefficients $A_{ij}$). Being of degree $s-1$, the second derivative of the collocation polynomial $\ddot{r}^\mu(\tau_n+\xi h)$ 
can be uniquely expanded in the Lagrange polynomials
\begin{equation}
L_i(\xi) = \prod_{k\neq i}^s\frac{\xi-c_k}{c_i-c_k}.\label{lagra}
\end{equation}
Setting $K_i^\mu = \ddot r^\mu(\tau_{n}+c_ih)$, this expansion can be written as
\[
\ddot r^\mu(\tau_n + \xi h) = \sum_{j=1}^sL_{j}(\xi)K_j^\mu,\quad \tau_n+\xi h\in[\tau_n,\; \tau_n+h].
\]
This implies that $\dot r^\mu$ and $r^\mu$ can be approximated  on $[\tau_n,\; \tau_n+h]$ by 
\begin{align}
\dot r^\mu(\tau_n+\xi h) &= u^\mu_n + \sum^s_{i=1}K_i^\mu A_i(\xi),& A_i(\xi)=& \int_0^\xi L_j(\xi')d\xi',\label{rkn12}\\
r^\mu(\tau_n +\xi h) &= x^\mu_n + \xi h \cdot u^\mu_n + \sum^s_{i=1}K_i^\mu B_i(\xi),& B_i(\xi)=& \int^\xi_0A_i(\xi')d\xi' = \int^\xi_0(\xi - \xi')L_i(\xi')d\xi'.\label{rkn13}
\end{align}
Now, by substituting $\xi = c_i$ for successive $i=1,...,s$, we can make the formulas (\ref{rkn11}), (\ref{rkn12}), (\ref{rkn13}) coinciding to Eqs.~(\ref{erk2_1}) - (\ref{erk2_3}), and thus specify the discrete coefficients of the collocation method:
\begin{align}
A_{ij} = A_j(c_i),\quad a_j=A_j(1),\quad B_{ij} = B_j(c_i),\quad b_j=B_j(1).
\label{iABab}
\end{align}
The forth order method can be obtained by using two Lagrange polynomials of degree one
\begin{align}
L_1(\xi) = \frac{\xi-c_2}{c_1-c_2},\quad 
L_2(\xi) = \frac{\xi - c_1}{c_2-c_1},
\end{align}
the resulting method is presented in Table~\ref{crkn}.
\begin{table*}[!ht]
 \caption{Double Butcher table for the collocation RKN method for $s=2$. 
    }\addtocounter{table}{-1}
   %%%% 
\subfloat[Collocation RKN]{
\label{crkn}
   \begin{tabular}[b]{c|c c} 
   $c_1$ &$\frac{c_1(c_1-2c_2)}{2(c_1-c_2)}$                            & $\frac{c_1^2}{2(c_1-c_2)}$ \\
   $c_2$&$\frac{c_2^2}{2(c_2-c_1)}$ &$\frac{c_2(c_2-2c_1)}{2(c_2-c_1)}$ \\\hline
   					        &$\frac{1-2c_2}{2(c_1-c_2)}$                             &$\frac{1-2c_1}{2(c_2-c_1)}$
   \end{tabular}
   \hspace{-2em}
   \begin{tabular}[b]{c|c c} 
    &$\frac{c_1^2(c_1-3c_2)}{6(c_1-c_2)}$                            & $\frac{c_1^3}{3(c_1-c_2)}$ \\
    &$\frac{c_2^3}{3(c_2-c_1)}$ &$\frac{c_2^2(c_2-3c_1)}{6(c_2-c_1)}$ \\\hline
   					        &$\frac{1-3c_2}{6(c_1-c_2)}$                             &$\frac{1-3c_1}{6(c_2-c_1)}$
   \end{tabular}
   }
\hspace{2em}
\subfloat[Gauss-Legendre RKN]{
\label{grk}
   \begin{tabular}[b]{c|c c|cc} %\hline
   $\frac{1}{2}-\frac{\sqrt{3}}{6}$ &$\frac{1}{4}$                            & $\frac{1}{4} -\frac{\sqrt{3}}{6}$ &$\frac{1}{36}$ &$\frac{5}{36}-\frac{\sqrt{3}}{12}$\\
   $\frac{1}{2}+\frac{\sqrt{3}}{6}$&$\frac{1}{4}+\frac{\sqrt{3}}{6}$ &$\frac{1}{4}$ & $\frac{5}{36}+\frac{\sqrt{3}}{12}$ &$\frac{1}{36}$\\\hline
   					        &$\frac{1}{2}$                             &$\frac{1}{2}$ &$\frac{1}{4} +\frac{\sqrt{3}}{12}$ & $\frac{1}{4} - \frac{\sqrt{3}}{12}$
   \end{tabular}
\label{GLRKN}
}\addtocounter{table}{1}
\end{table*}
The highest possible order $2s$ can be achieved when the collocation points are chosen as roots of the $s$th shifted Legendre polynomial \cite{Butcher:2008}
\begin{align}
\frac{d^s}{d\tau^s}\left[\tau^s(\tau-1)^s\right] = 0,\quad \tau\in (0,\;1).
\label{lg}
\end{align}
Other possible choices of the collocation points (Gauss-Lobatto, Gauss-Radau methods) possess lower order $p < 2s$, but might be still beneficial if e.g. the equations are stiff \cite{Norsett:II}. For $s=2$ the roots of the quadratic polynomial~(\ref{lg}) are $c_{1,2} = 1/2 \mp \sqrt{3}/6$ and the resulting Gauss-Legendre-RKN method of order $4$ is  presented in Table~\ref{grk}. 

In this paper we also use higher order methods based on $s=3$ and $s=4$ collocation points. 
The Gauss-Legendre-RKN method with $s=3$ collocation points is briefly described in Appendix~\ref{irkn6}. Though  computationally  more expensive, these $6$th and $8$th order methods provide possibility to measure the accuracy of lower order methods in case analytical solution is not available.

After the coefficients~(\ref{iABab})  are substituted into the system (\ref{erk2_1})-(\ref{erk2_3}) it turns to a set of stage equations. Implicitness of the collocation RK(N) methods implies usage of the iteration technique to solve the stage equations. The simplest iteration scheme can be obtained by just hanging  the iteration induces $m$, $m+1$ on the implicit equations:
\begin{alignat}{2}
\mbox{iRK(N):}\quad K_i^{\mu,m+1} &= 
f^\mu(\tau_n+c_ih,\; x^\mu_n +c_ih u^\mu_n+ h^2\sum^s_{j=1}B_{ij}K_j^{\mu,m},\; u^\mu_n + h\sum^s_{j=1}A_{ij}K_j^{\mu,m}),
\end{alignat}
where the prefix ``i'' stands for ``implicit'' to distinguish the method from explicit ones, for the latter we will similarly use the prefix ``e'' in front of their abbreviation RK(N). In order to formulate the criterion for stopping the iteration process, an error of the approximation has to be measured by an appropriate norm. In this paper we use for this purpose the maximum norm 
\begin{align}
||K^{m+1}-K^{m}||_\infty = \max_{i,\mu}| K_i^{\mu,m+1}-K_i^{\mu,m}|.
\end{align}
In the rest of the paper, we discuss applications of the described numerical methods to physical problems of particle motion in ultrastrong electromagnetic fields.

\section{Particle in magnetic field}
\label{mag_test}

\begin{figure}[ht]
\centering
  \subfloat[Particle trajectories.]
  {\label{mtraj1}\includegraphics[width=0.48\textwidth]{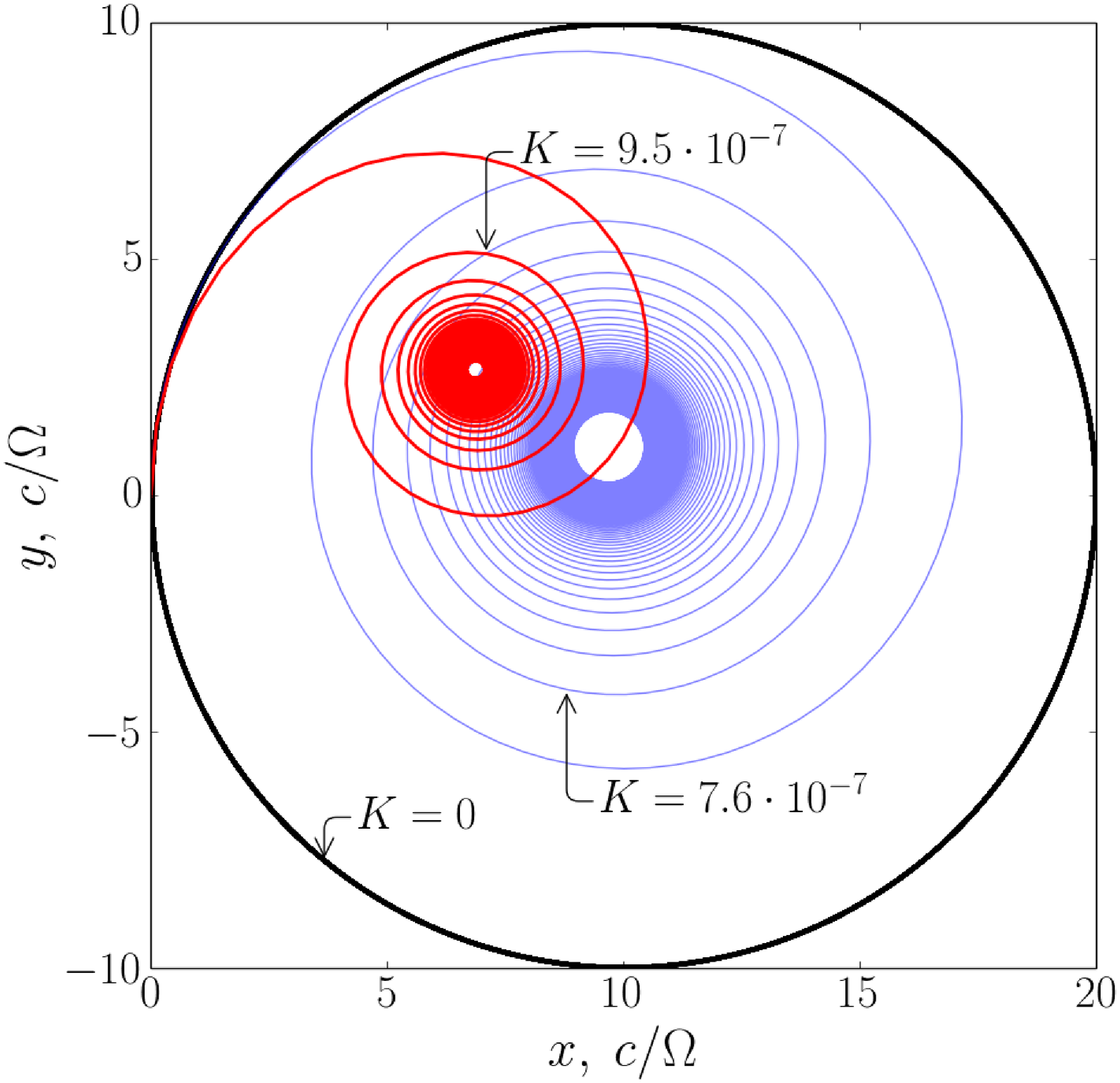}} 
  \hfill        
\subfloat[On top: dependence of the error measure $L_2$ upon a time step; on bottom: inaccuracy in preservation of the on-shell condition.]
  {\label{mtraj2}
  \includegraphics[width=0.48\textwidth]{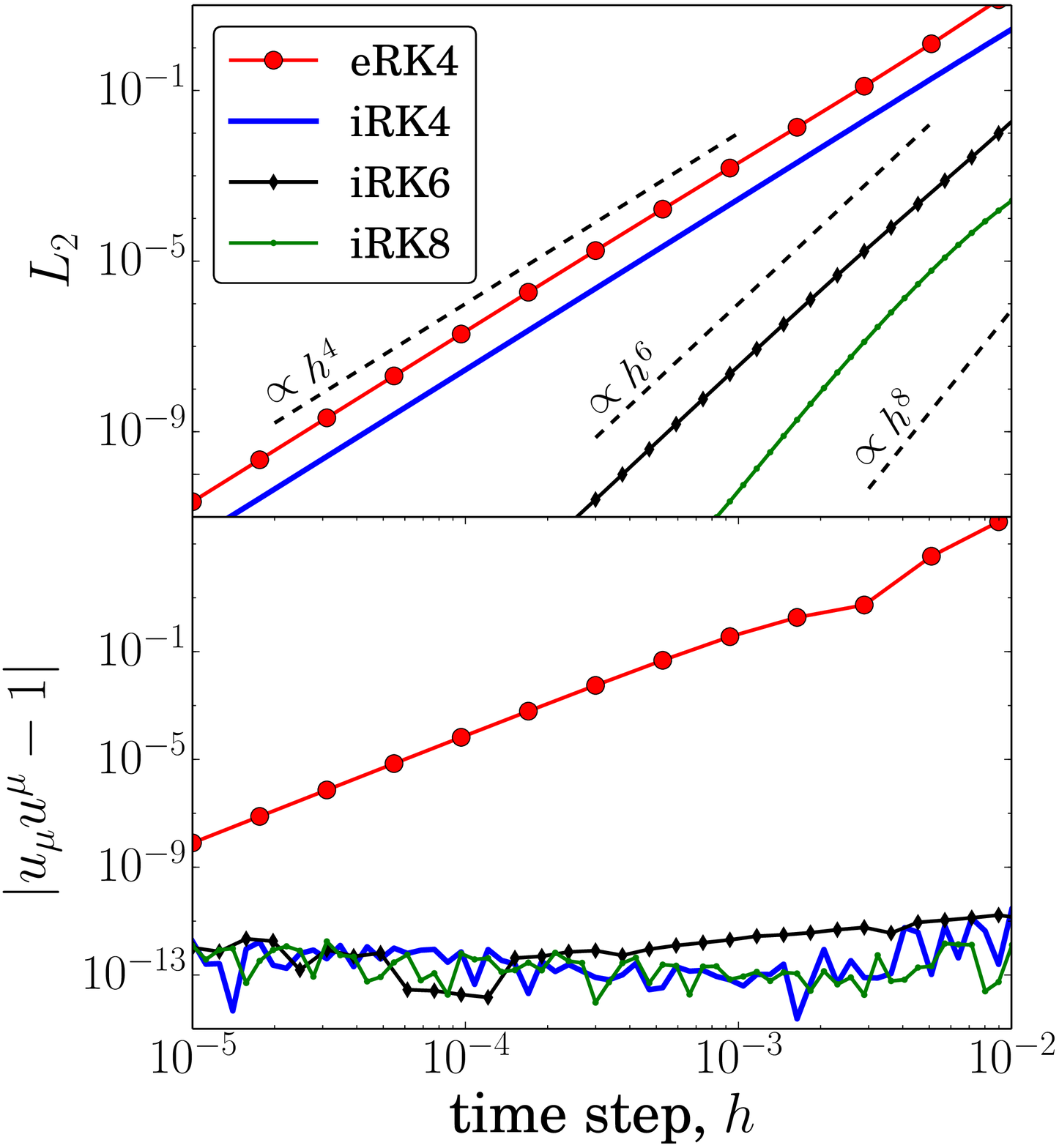}
  }
\caption{(Color online). Effect of the radiation reaction on particle motion in a constant magnetic field $eH_z/m\Omega = 10^3$. Initial $\gamma$-factor of a particle is $\gamma_0 = 10^3$.}  
\label{mtraj}
\end{figure}

Let us start with motion of an ultrarelativistic particle in a constant magnetic field. By fixing the direction of magnetic field along $x^3 = z$, the LL equation reads 
\begin{align}
\frac{d\vec u_\perp}{d\tau} &= \frac{e}{m}\vec u_\perp\times\vec H - \frac{2}{3}\frac{e^4}{m^3}H^2(1 + u^2_{\perp})\vec u_\perp,\label{magtau1}\\
\frac{du_z}{d\tau} &= -\frac{2}{3}\frac{e^4}{m^3}H^2u^2_\perp u_z,
\label{magtau2}
\end{align}
where $\vec u_\perp = \{u_x,\; u_y\}$ is the transverse component of 4-velocity. After passing to polar coordinates $u_x = u_\perp\cos\phi$, $u_y = u_\perp\sin\phi$ in the transverse plane $xy$ and introduction of dimensionless time $\varphi = \Omega\tau$ (where $\Omega = eH/m$ is the rotation frequency in the proper reference frame), the system~(\ref{magtau1}) - (\ref{magtau2}) takes  the form 
\begin{align}
\frac{du_\perp}{d\varphi} &= -Ku_\perp(1+u^2_\perp)\label{magphi1}\\
\frac{d\phi}{d\varphi} &= -\sigma,\label{magphi2}\\
\frac{du_z}{d\varphi} &= -Ku^2_\perp u_z.
\label{magphi3}
\end{align}
Here $\sigma=\pm 1$ is the sign of a particle charge and $K$ is the dimensionless parameter
\begin{align}
K = \frac{2}{3}\frac{e^2\Omega}{m} = \frac{2}{3}\frac{e^3H}{m^2} = \frac{2}{3}\frac{\alpha H}{H_c},
\end{align}
where $H_c = m^2/e$ is the critical magnetic field. Note that under applicability of LL equation $K\ll 1$. Eq.~(\ref{magphi1}) admits separation of variables, after that the rest equations (\ref{magphi2}) - (\ref{magphi3}) can be also easily solved. The solution reads
\begin{align}
\vec{u}(\varphi) = \frac{1}{\sqrt{\left(1+u_{\perp 0}^2\right)e^{2K\varphi}-u^2_{\perp 0}}}
\left\{u_{\perp 0}\cos(\phi_0 -\sigma\varphi),\; u_{\perp 0}\sin(\phi_0-\sigma\varphi),\; u_{z0} e^{K\varphi}\right\},
\label{exmag}
\end{align}
where the subscript ``0'' refers to the initial values at $\varphi=0$. By simple integration
one can find the particle trajectory in terms of hypergeometric function, e.g. its projection onto the plane $xy$ is given by
\begin{align}
\left[\begin{array}{c}
x-x_0\\
y-y_0
\end{array}
\right] = \frac{\sigma}{u_{\perp0}}\left\{
\sqrt{\left(1+u_{\perp 0}^2\right)e^{2K\varphi} - u^2_{\perp0}}
\left[\begin{array}{c}
\Im\\
-\Re
\end{array}
\right]
e^{i(\phi_0-\sigma\varphi)}\cdot {}_{2}{F_1}
\left(1,\;\frac{1}{2}-\frac{i\sigma}{2K},\; 1-\frac{i\sigma}{2K},\; \frac{(1+u^2_{\perp0})}{u^2_{\perp0}}e^{2K\varphi}\right)
\right\}\Bigg|^\varphi_0,
\end{align}
(below we assume $u_{z0} = 0$). The trajectory is a converging spiral shown in Fig.~\ref{mtraj1}. Trajectory without radiation reaction corresponds to the case $K=0$ and represents a circle. Numerical integration is performed with the RK method, since this particular problem does not contain dependence on coordinates. The observed order of accuracy of the method can be estimated by examining the behaviour of $L_2$ norm 
\begin{align}
L_2 = \sqrt{\frac{h}{T}\sum_{i}|u^\mu(\tau_i) - u^\mu_{\rm ex}(\tau_i)|^2},
\end{align} 
where $T$ is the proper time duration of simulation and $u^\mu_{\rm ex}$ refers to the exact solution~(\ref{exmag}). 

The results of studies of convergence of different numerical methods are presented in Fig.~\ref{mtraj2}, where both $L_2$ and the Minkowski norm errors are plotted in dependence upon a time step $h$. The slopes of the curves in the upper panel of Fig.~\ref{mtraj2}, corresponding to both the explicit eRK4 and implicit iRK4 methods clearly correspond to the order four, as expected, but the difference in the initial error level makes the implicit method more favourable with respect to higher accuracy. The higher order methods iRK6 and iRK8 also exhibit the expected slopes, corresponding to orders six and eight, respectively. The inaccuracy of preservation of the mass shell condition, as shown in the lower panel of Fig.~\ref{mtraj2}, remains low and bounded for all implicit methods, while the eRK method results in  substantial violation of the Minkowski norm. It should be noted that non-conservation of the Minkowski norm depends on the numerical error $\delta u^\mu$ as
\[(u^\mu+\delta u^\mu)(u_\mu+\delta u_\mu)=1+2u^\mu\delta u_\mu+O(\delta u^2),\]
and hence is especially pronounced in ultrarelativistic case $u^\mu\gg 1$.
 
\section{Particle in plane wave field}
\label{pw_test}

\begin{figure}[ht]
\centering
  \subfloat[Circular polarization]
  {\label{traj_lin1}\includegraphics[width=0.48\textwidth]{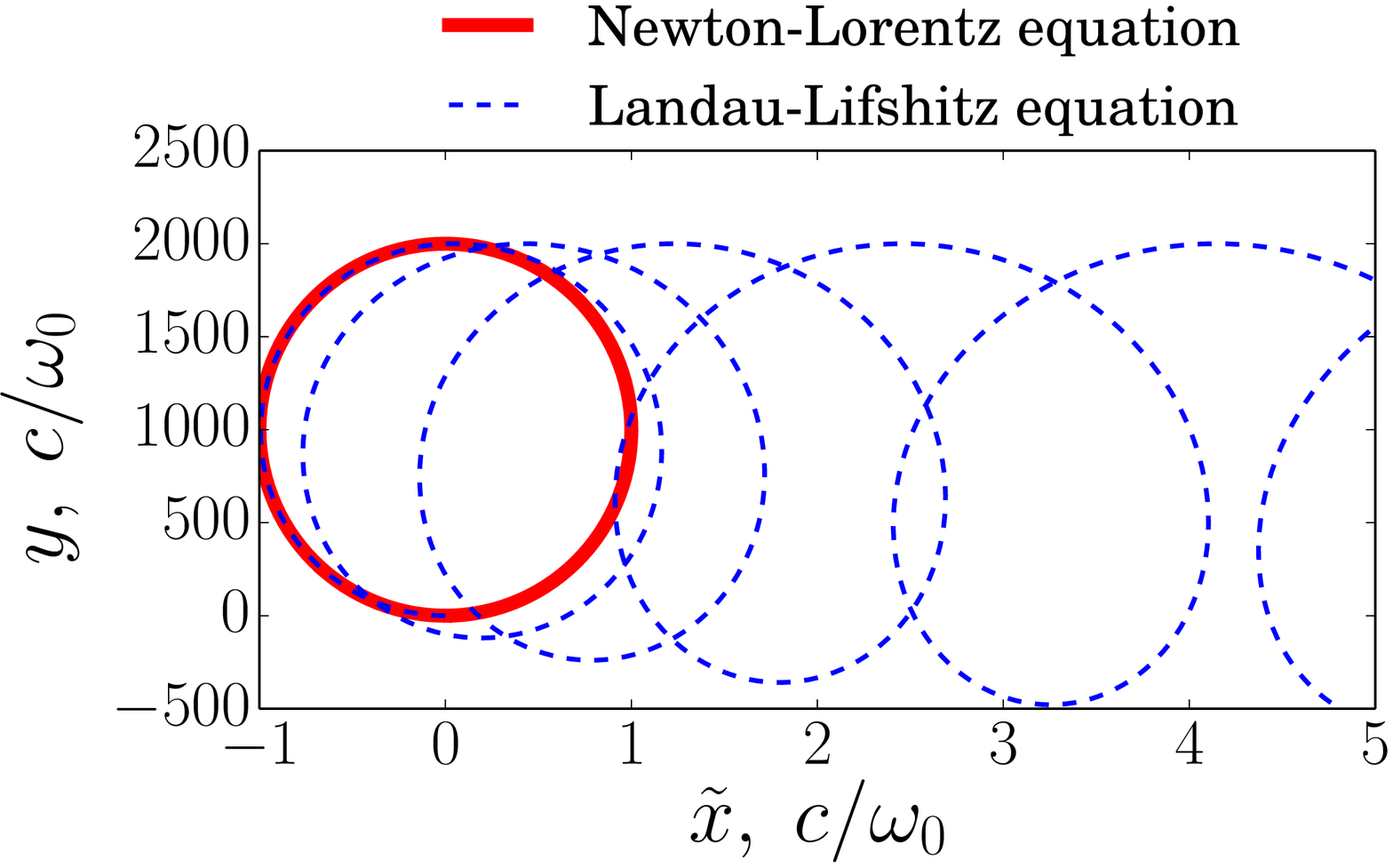}}
  \hfill
  \subfloat[Linear polarisation]
  {\label{traj_lin2}\includegraphics[width=0.48\textwidth]{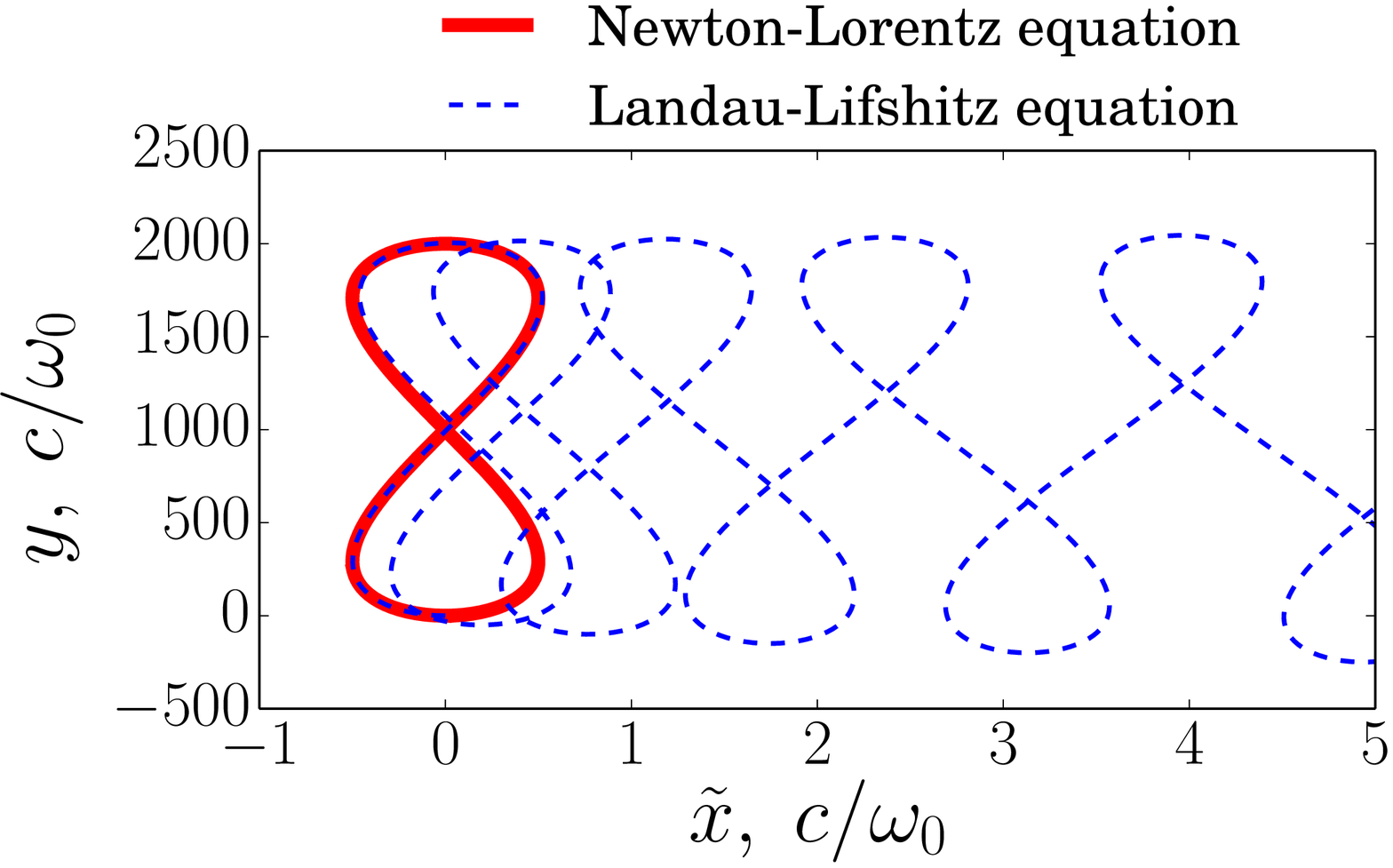}}
 % \hfill
\caption{(Color online). Trajectories of a charge in a plane wave field for $a_0=10^3$ in a frame that would be comoving for a guiding center in the absence of radiation reaction. Radiation reaction leads to a drift of guiding center along the wave vector.} \label{traj_lin}
\end{figure}  

The LL equation admits much simpler form for a particle moving in a plane wave field $A_\mu(x) =  A_\mu(\varphi)$, $\varphi =k x= \omega_0 t - \vec k\cdot\vec r$, $k A= 0$:
\begin{align}
\frac{du^\mu}{d\varphi} = \frac{e}{m}\left[h (A'u) k^\mu - {A^\mu}'\right]+\frac{2}{3}\frac{e^3}{m^2 h}\left[h(A''u) k^\mu-{A^\mu}''\right]-\frac{2}{3}\frac{e^4}{m^3}{A'}^2 k^\mu+\frac{2}{3}\frac{e^4}{m^3h}{A'}^2 u^\mu
\end{align}
where $1/h=\dot\varphi = k_\mu u^\mu$. This equation can be  solved analytically in quadratures \cite{diPiazza:2008} and in special cases the 4-velocity $u^\mu$ can be even expressed in terms of elementary functions \cite{Hadad:2010}. Here we consider two important cases corresponding to linear and circular polarisation of the wave. 

The effect of radiation reaction on a particle trajectory is particularly transparent in a reference frame comoving with the guiding center, as in this frame the radiation-free trajectories take especially simple form: in case of linear polarization particles move along the figure-8, whereas in case of circular polarization the trajectory is a circle. The radiation reaction effects open closed figure-8 or circular motion and lead to a drift along the wave vector of the wave (see Fig.~\ref{traj_lin}). 

\begin{figure}[ht]
\centering
  \subfloat[Convergence in $L_2$ norm]
  {\label{traj_circ1}
   \includegraphics[width=0.48\textwidth]{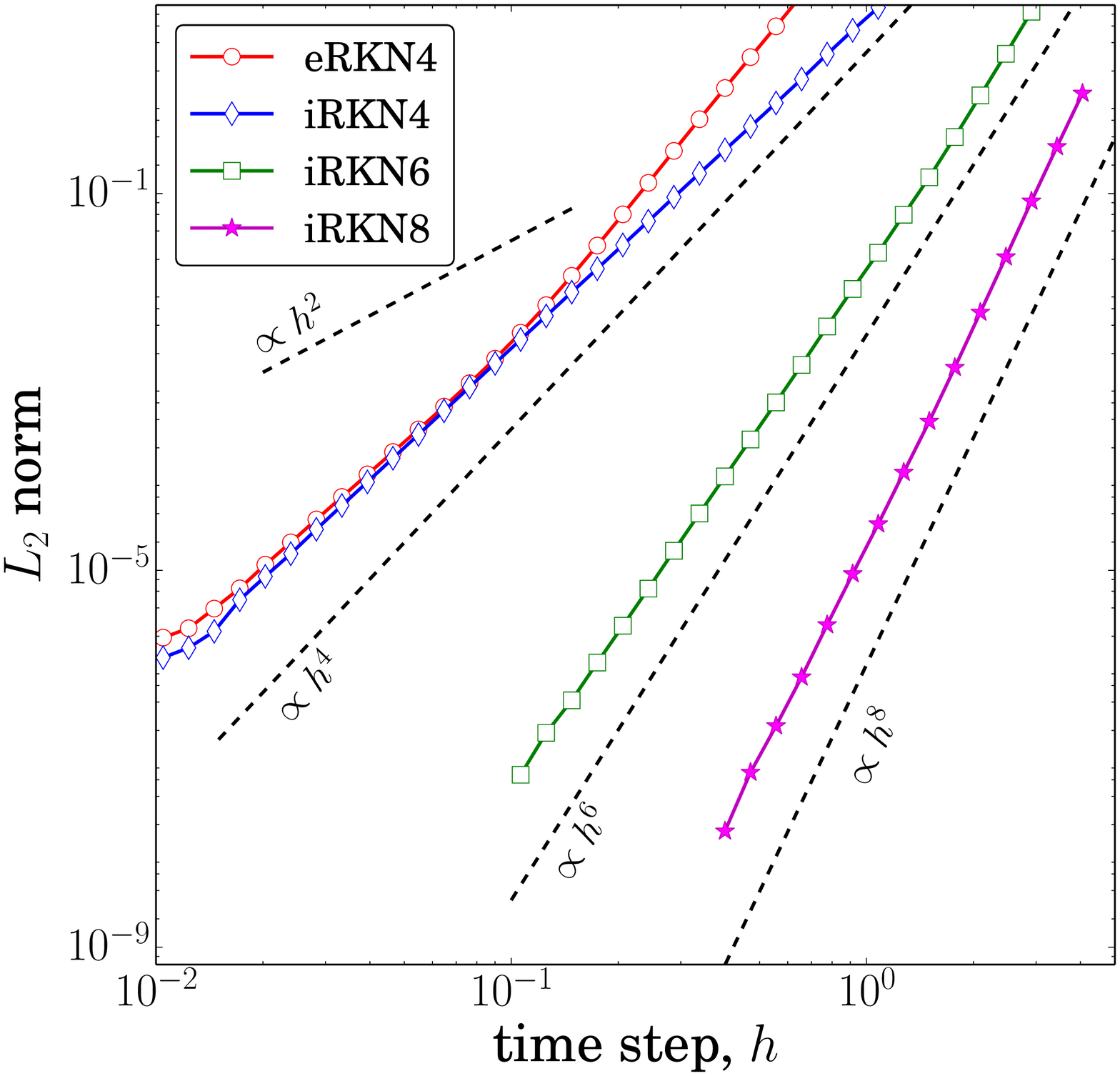}
  }
  \hfill
  \subfloat[Error in Minkowski norm]
  {\label{traj_circ2}
   \includegraphics[width=0.48\textwidth]{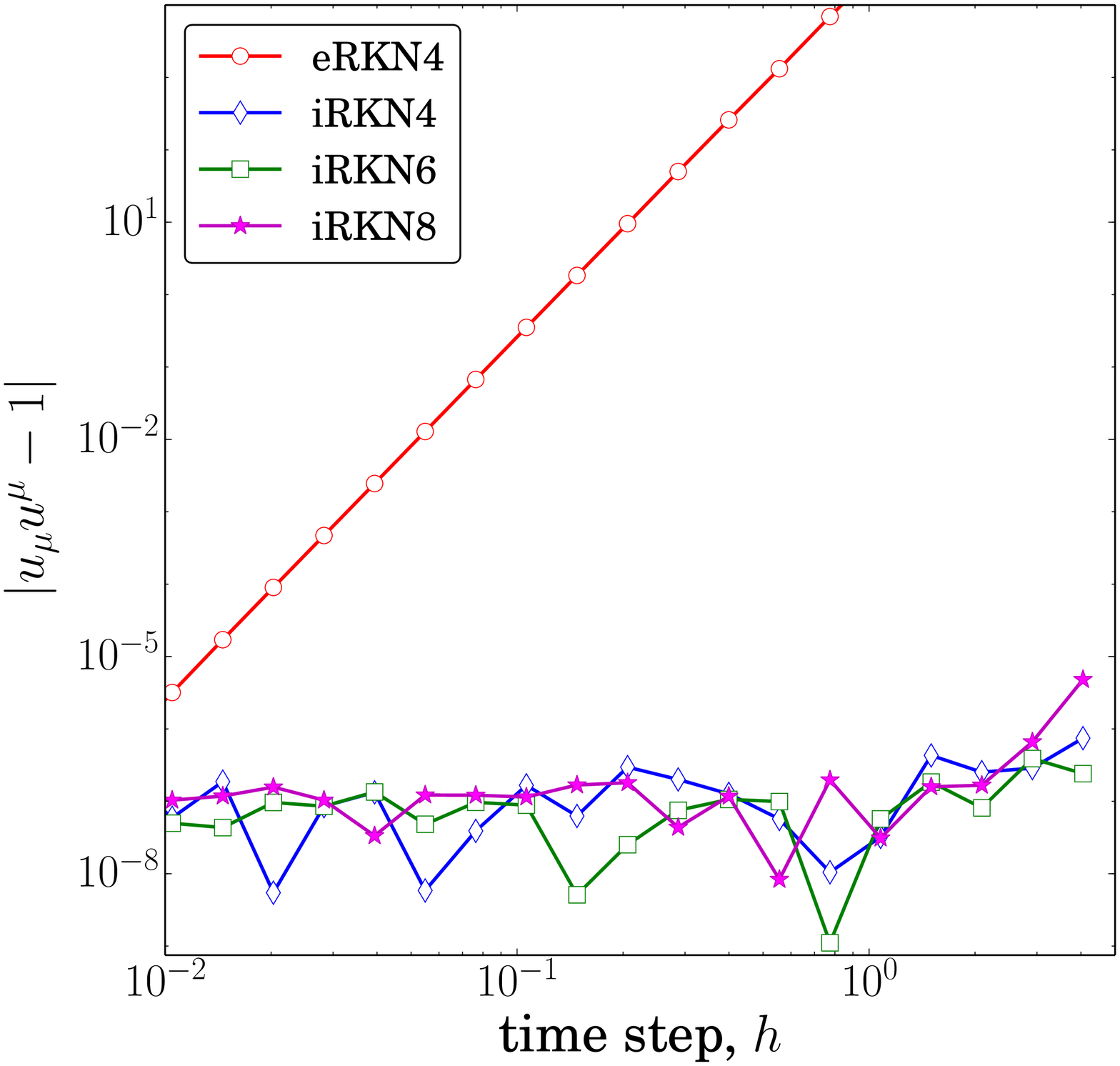}
  }
\caption{(Color online). Convergence study for a particle moving in a circularly polarised plane wave for $a_0 = 10^3$.}  
\label{traj_circ}
\end{figure}

In order to gain more knowledge about the accuracy of the numerical methods we compare numerical and 
analytical solutions in case of a circularly polarized plane wave, for which the dependence of 
the phase $\varphi$ and therefore of $u^\mu$ and $ x^\mu$ on the proper time $\tau$ can be expressed explicitly by \cite{Hadad:2010}
\begin{align}
\phi(\tau) = \frac{\sqrt{1+2a_0^2(ku_0)^2K\tau/\omega_0}-1}{a^2_0(ku_0)K/\omega_0},
\label{phi_tau}
\end{align}
where $K$ is this time defined by $K=(2/3)e^2\omega_0/m$. The results are presented in Fig.~\ref{traj_circ}. One can see that the slopes of the curves in Fig.~\ref{traj_circ1} are in agreement with the expected orders of the methods, indicated by dashed lines. Only the curve for eRKN4 method on this plot deviates slightly from power law when the time step becomes larger. As for Fig.~\ref{traj_circ2}, it demonstrates the same behavior of the methods as was previously pointed out for the problem of motion in a constant magnetic field (compare to Fig.~\ref{mtraj2}).

\begin{figure}[ht]
\centering
   \includegraphics[width=0.48\textwidth]{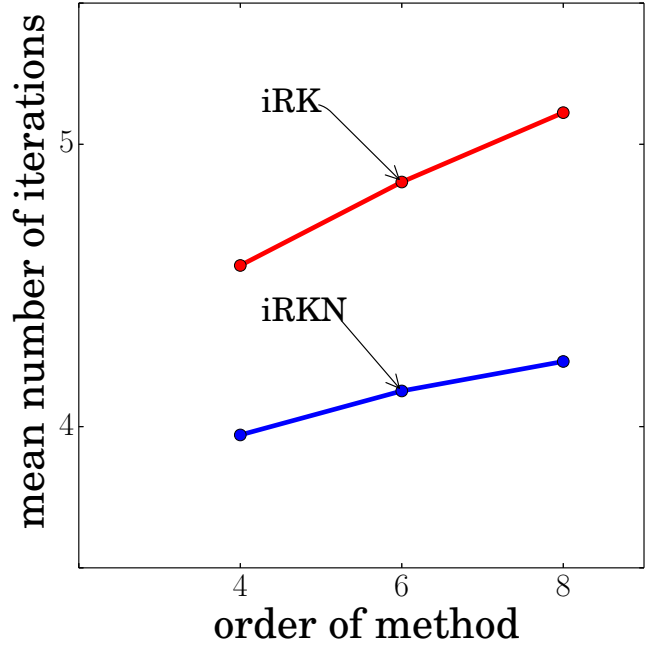}
\caption{(Color online). Computational complexity for a particle moving in a linearly polarized plane wave for $a_0 = 10^3$.}  
\label{figure_6}
\end{figure}

Let us discuss computational complexity of the implicit methods. The main source of computational complexity in implicit methods comes from the number of iterations needed for a nonlinear solver to achieve the desired accuracy. Hence we measure it for both the RK and RKN methods by an average number of iterations per step. The results for a case of motion in the field of linearly polarized plane wave are summarized in Fig.~\ref{figure_6}. 

According to Fig.~\ref{figure_6}, the iRKN method requires less iterations than iRK. One can also compare  qualitatively the performance of implicit iRKN4 and its explicit counterpart eRKN4. The number of stage equations (\ref{erk2_1}) for eRKN4 equals $4$, while the corresponding number of stages for iRKN4 is $2$.  But because of $4$ iterations on the average for the same time step, the iRKN4 method is twice more expensive in the number of calls of the right-hand-side function. From the other hand the implicit methods admit longer time steps without deterioration of numerical accuracy. Moreover, further increase of the order of explicit methods eventually results in reaching the so-called Butcher barrier \cite{Butcher:2008}, after which the amount of stages starts to grow faster than the order.  
 
\section{Scattering of particle on focused laser pulses}
\label{pond}
In this section we apply the developed numerical technique to an important problem of scattering of particles by focused laser pulses. In the absence of radiation reaction, this problem can be studied with the theory of ponderomotive scattering, see e.g. \cite{Bauer:1995,Naro:2000}. In previous studies of this problem radiation reaction was neglected because of the assumption of relatively small both the intensity  and the particle energy. Our basic settings are  similar to those considered in \cite{Naro:2000} but with the laser intensity and particle energy higher by two order of magnitude, $\gamma_0,\,a_0\sim 10^3$. Consider first a focused monochromatic Gaussian beam with the electric field defined by
\begin{align}
\vec E = \frac{iE_0b^2}{(b + i\vec r\cdot\vec n)^2}
\left\{\vec \varepsilon + \frac{\omega [\vec r\cdot(\vec \varepsilon\times\vec n)]
(\vec r\times\vec n)}{b+i\vec r\cdot\vec n}\right\}e^{-i\omega_0(t-\vec r\cdot\vec n)}
\exp\left[-\frac{\omega(\vec r\times\vec n)^2}{2(b+i\vec r\cdot\vec n)}\right].
\end{align}
Our notations here are the same as in Ref.~\cite{Fedotov:2009}: $\omega_0$ is the carrier frequency of the beam, $b$ is a half of the Rayleigh length, $\vec \varepsilon$ is a (complex) polarization vector, and $\vec{n}$ is a unit vector along the focal axis. The magnetic field is given by $\vec H = -(ic/\omega_0)\nabla\times\vec E$. Our target configuration consists of two beams with the same polarization $\vec \varepsilon$, but oppositely directed vectors $\vec n$. In such a case near the focal axis the electric field is doubled, while the magnetic field vanishes. However, in the rest of the focal plane $\vec n\cdot\vec r = 0$ only a component $\vec H_\perp$ in this plane is vanishing, while its longitudinal component $\vec H\parallel \vec n$ is also doubled. Nevertheless, the component $\vec H_\parallel$ is of the order of $O(\Delta)$, where $\Delta=(2\omega_0 b)^{-1/2}$ is the angular aperture of the beams, assumed to be sufficiently small. Below, we consider a particular case of circular polarization of the beams in a sense discussed in Refs.~\cite{Naro:2000,Fedotov:2009}. 

\begin{figure}[ht]
\centering
  \subfloat[Scattering on laser focus]{\label{B1}\includegraphics[width=0.48\textwidth]{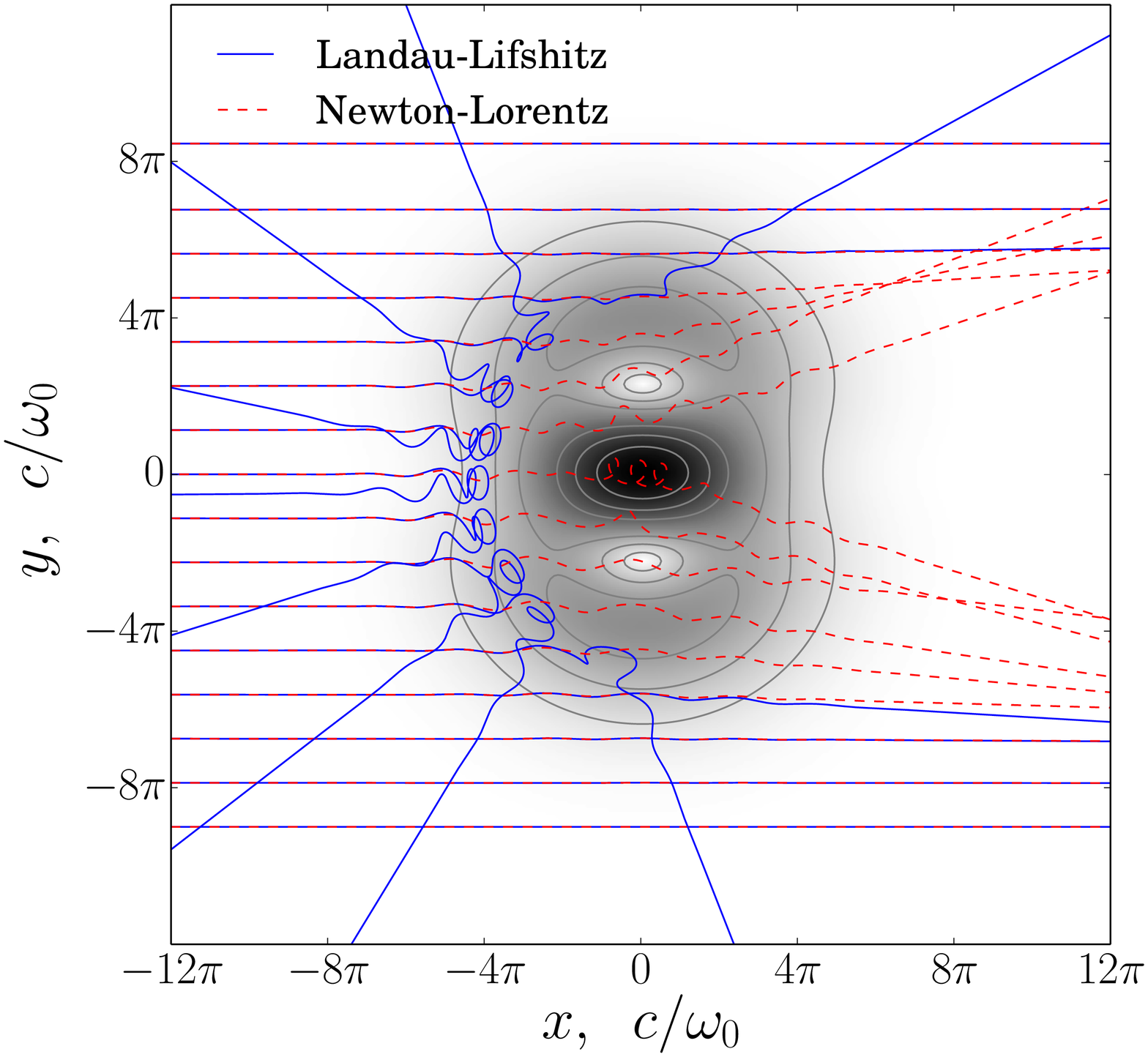}} 
  \hfill        
  \subfloat[Energy spectra after scattering]{ \label{sf2}\raisebox{2mm}{
  \includegraphics[width=0.48\textwidth]{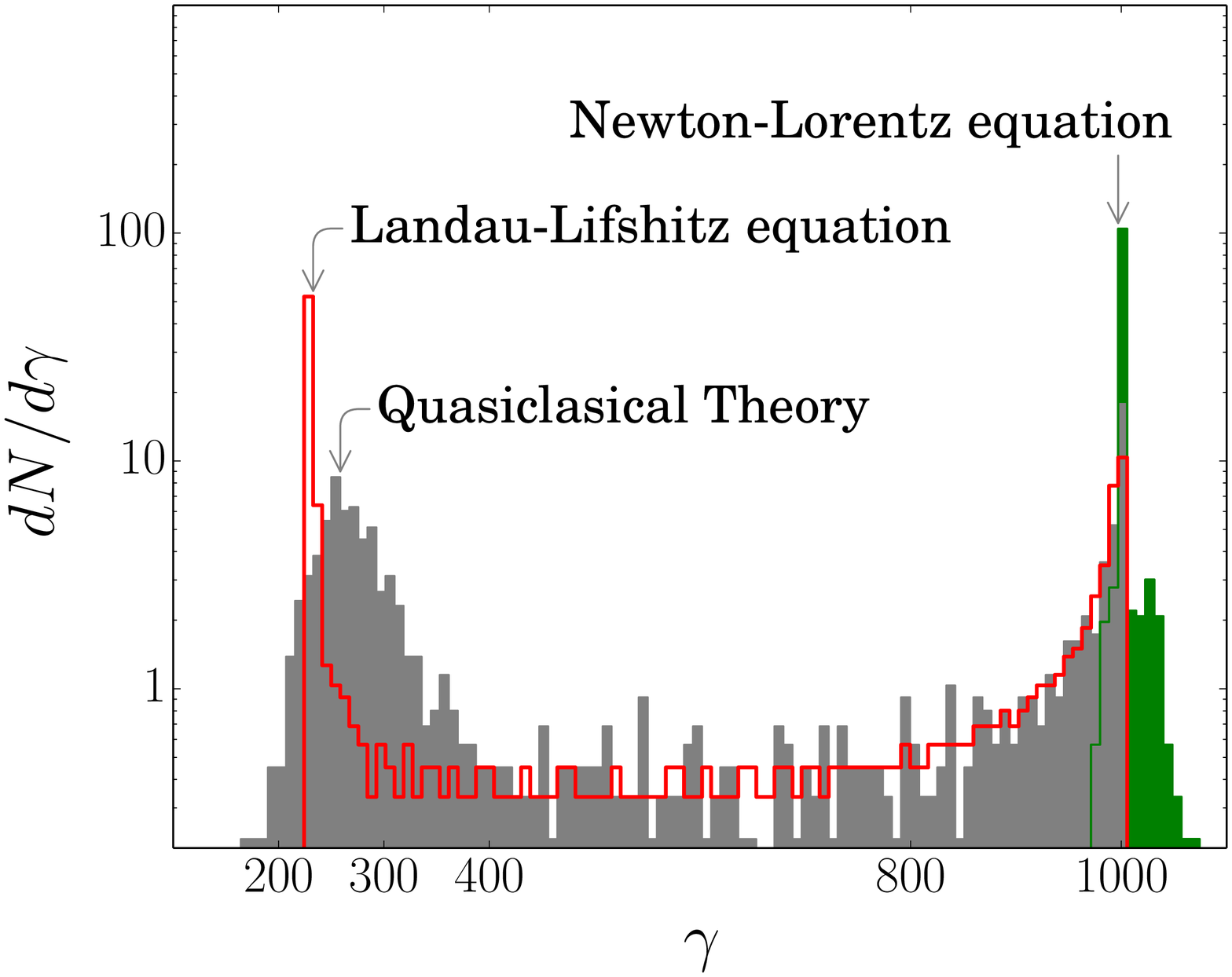}}}
  %%%
\caption{(Color online). \protect\subref{B1} Trajectories of scattered particles with radiation reaction (solid blue lines), and without radiation reaction (dashed red lines). \protect\subref{sf2} Energy spectrum of particles after they have left the focal region. Initial Lorentz factor is $\gamma_0 = 10^3$ and dimensionless field amplitude is $a_0 = 10^3$, focusing parameter is $\Delta=0.1$ ($b=31.8\lambda$).}\label{scatt}
\end{figure}

\begin{figure}[ht]
\centering
   \includegraphics[width=0.48\textwidth]{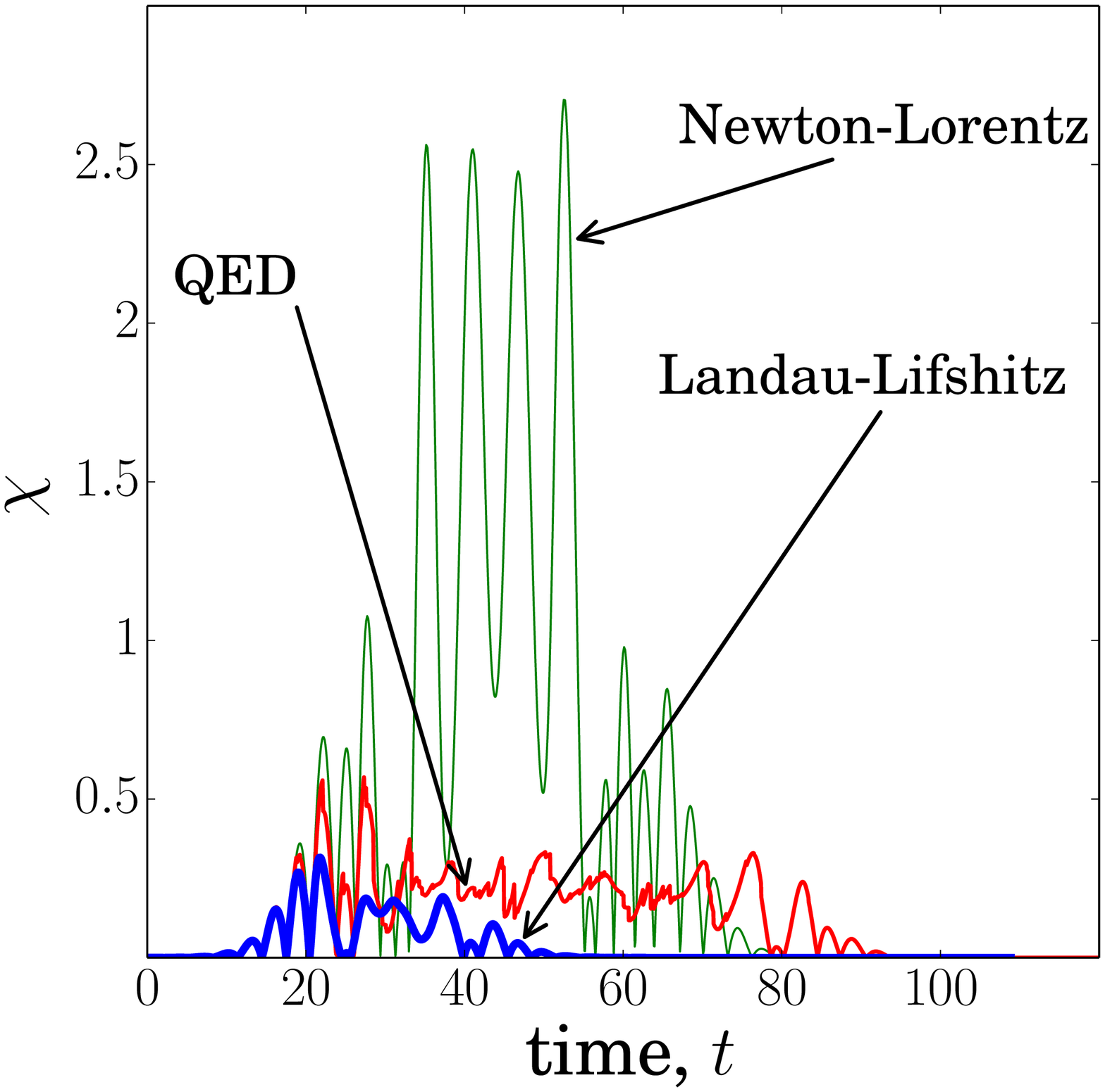}
\caption{(Color online). Quantum dynamical parameter $\chi$ as a function of time 
for a particle with impact parameter zero.}  
\label{figure_7}
\end{figure}

The ponderomotive potential can be defined as follows \cite{Bauer:1995,Naro:2000}
\begin{align}
U(\vec r) = m\sqrt{1 +\frac{e^2}{m^2\omega_0^2}|\vec E_{tot}(\vec{r},t)|^2},
\label{upot}
\end{align} 
where $\vec E_{tot} = \vec E_{\vec n} + \vec E_{-\vec n}$ is the total electric field of counterpropagating beams. The scattering picture shown in Fig.~\ref{scatt} was calculated for several electrons with different impact parameters $\rho$ and initial Lorentz factor $\gamma_0 = 10^3$, for $\vec{n}$ directed along the axis $z$, $\vec{\varepsilon}=\{1,i,0\}$ and the initial momenta of all the electrons  directed along the axis $x$.  
\begin{figure}[ht]
\centering
  \subfloat[Scattering angle $\theta$]{\label{b1}
  \includegraphics[scale=0.3]{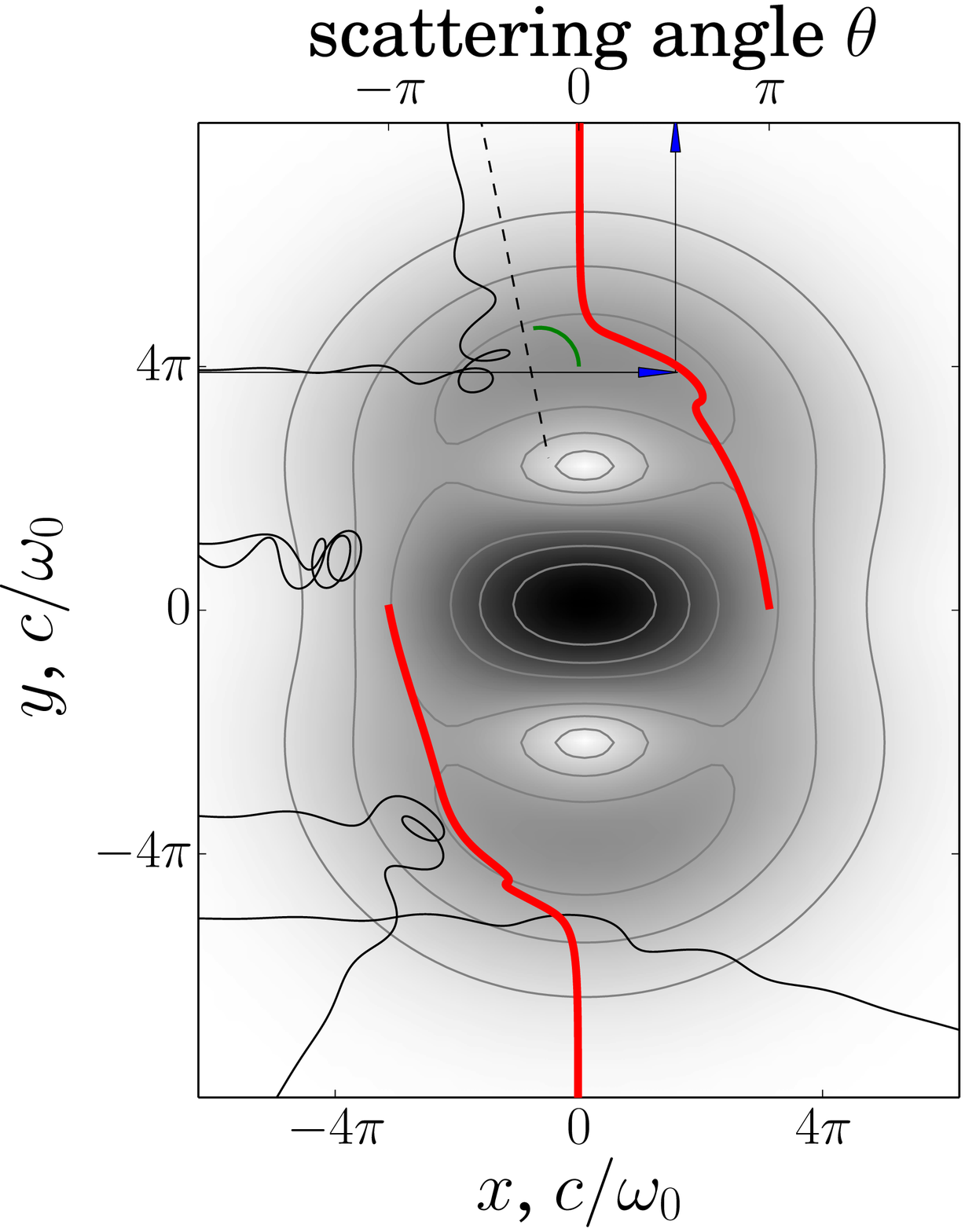}} 
  \subfloat[Relative angular error]{\label{b2}
  \includegraphics[scale=0.3]{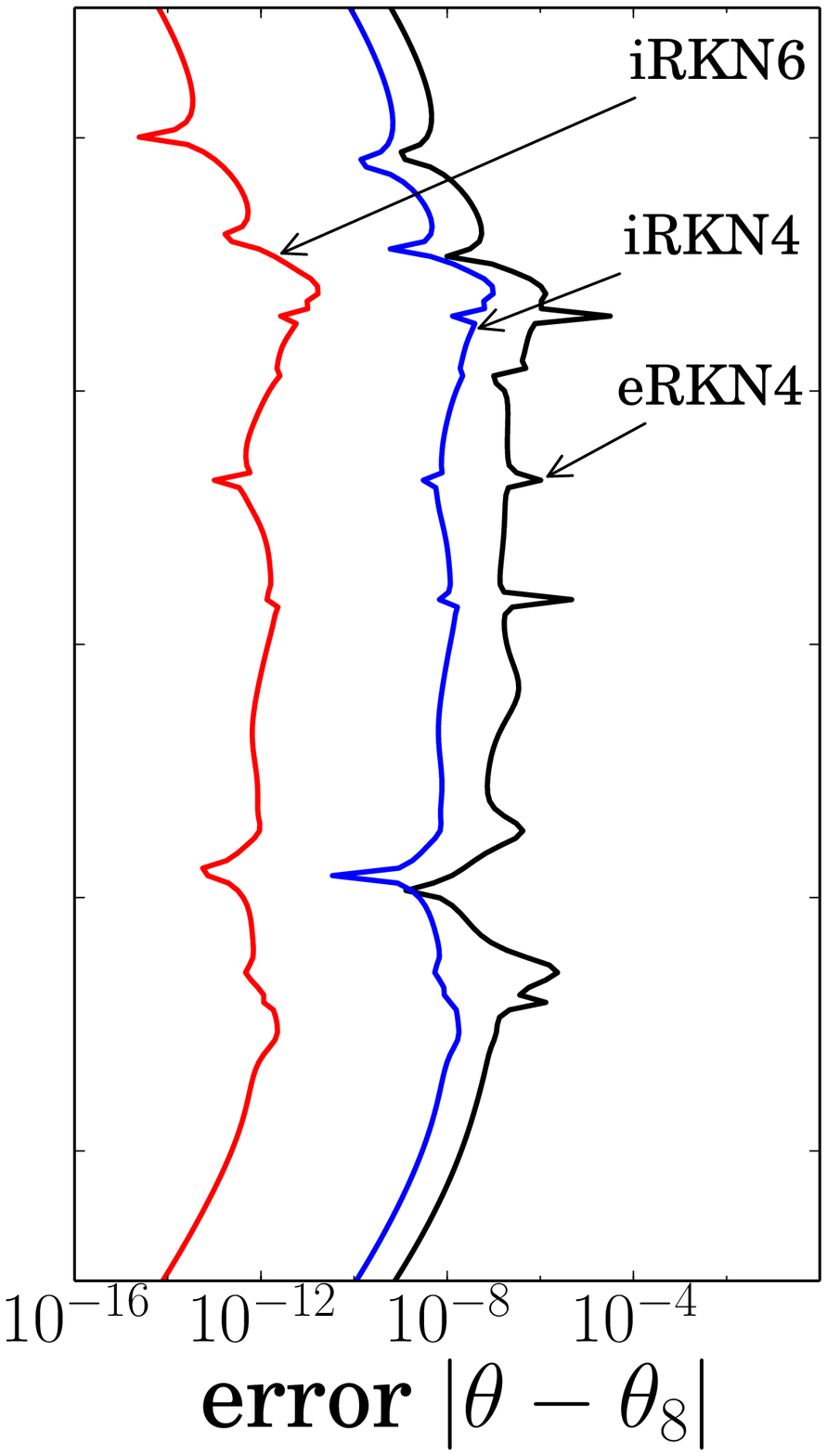}} 
%  \hfill        
  \subfloat[Minkowski norm error]{\label{b3}
  \includegraphics[scale=0.3]{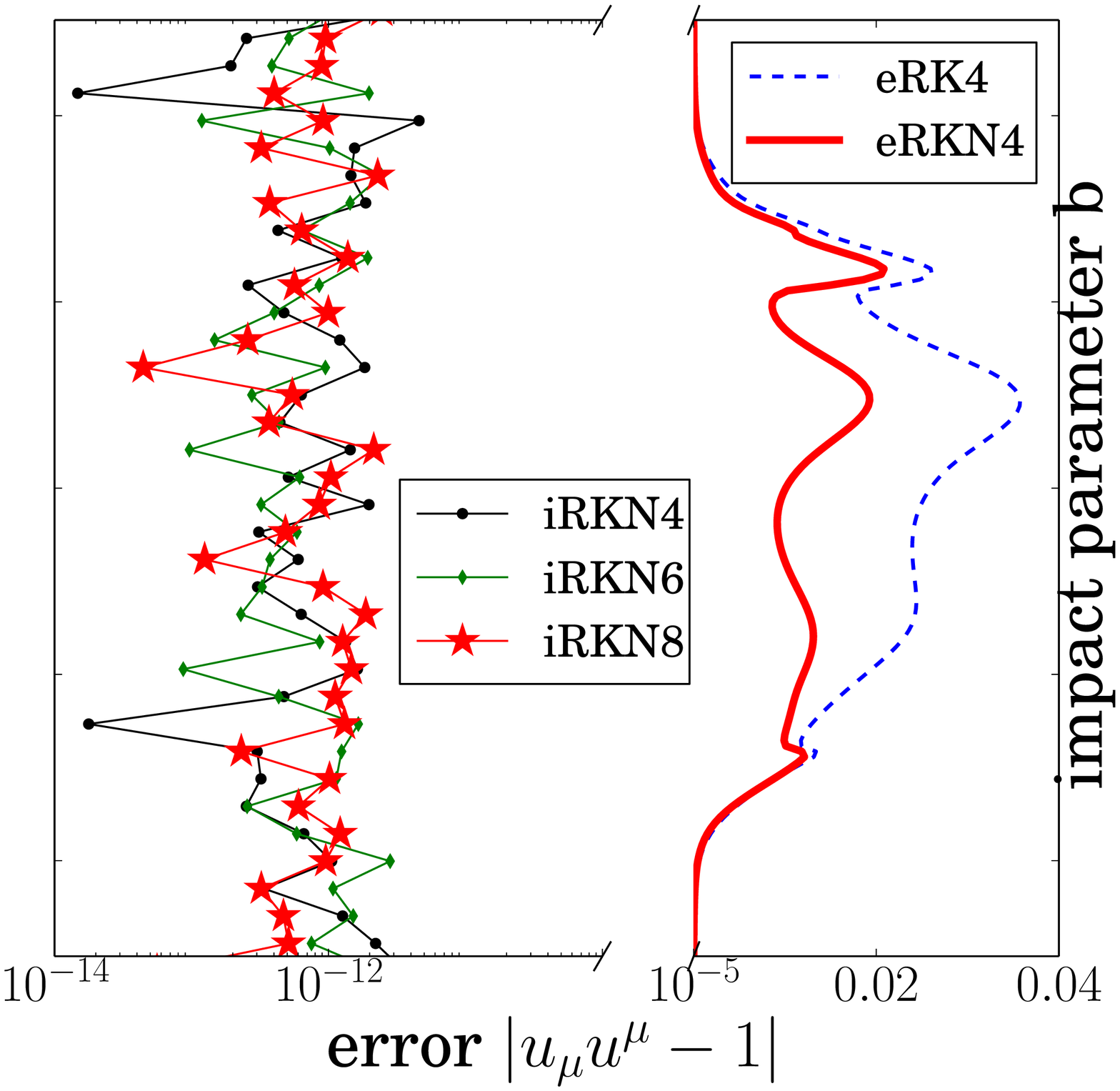}
  }
\caption{(Color online). \protect\subref{b1} Dependence of the scattering able on the impact parameter; \protect\subref{b2} and \protect\subref{b3} estimations of accuracy for numerical simulation of the scattering problem using the relative scattering angle error and the Minkowski norm (parameters are the same as in Fig.~\protect\ref{scatt}). Simulation is performed with the time step value $h=10^{-4}\omega_0^{-1}$.} \label{scatt1}
\end{figure}

An estimate for the particle energy (initial Lorentz factor $\gamma_0$) required to overcome the ponderomotive potential barrier without taking account for radiation reaction was formulated in \cite{Naro:2000} 
\begin{align}
\gamma_0 >\sqrt{1 + a_0^2}.
\label{reflect}
\end{align}
Even though this criterion was derived only under some restrictions, e.g. for head on collision with a single beam possessing azimuthal symmetry of ponderomotive potential, and need not be recognized literally in our setup, it can be still used as a useful estimate for tuning the initial conditions close to the threshold values.

Unlike previous sections, the problem under consideration does not admit analytical solutions, so that it can be explored by numerical simulations only. The results of simulation of scattering of electrons incident in the focal plane transversely to the counterpropagating beams both with and without taking account for radiation reaction are presented in Fig.~\ref{scatt}. If the scattering is described by the NL equation (Fig.~\ref{B1}, red trajectories), then all the electrons with $\gamma_0=10^3$ cross the focus without reflection. However, the radiation reaction effects lead to total reflection of the electrons from the laser focus (blue trajectories).  

The energy spectra after scattering are depicted in Fig.~\ref{sf2}. The energy distribution obtained from the NL equation is concentrated around the initial energy corresponding to $\gamma_0 = 10^3$, with a small dispersion $\Delta\gamma \sim 50$ (corresponding to deviations from the ponderomotive theory). The spectrum obtained with the LL equation is considerably shifted towards the lower energy and is peaked at $\gamma_1^{(cl)} \simeq 220$. 
For the sake of completeness we supply this plot with the results obtained with the quantum theory of radiation \cite{Elkina:2011}. One can see that the quantum picture results in a wider energy spectrum peaked at a bit higher value $\gamma_1^{(q)}\simeq 250$. In order to explain why the results obtained with LL equation and with quantum approach deviate only slightly, in spite of that the explored value of $a_0$ is high, let us consider applicability of LL equation in more details. As is well known, quantum and classical radiation regimes are discriminated  by the quantum dynamical parameter \cite{Nikishov:1967}
\begin{align}
\chi = \frac{e}{m^2}\sqrt{-(F^{\mu\nu}u_\nu)^2} = 
\frac{e}{m^2}\sqrt{\left(\gamma\vec E + \vec u\times\vec H\right)^2-(\vec u\cdot\vec E)^2}.
\label{chi}
\end{align}
Specifically, if $\chi\ll 1$, then radiation reaction is classical, otherwise it must be treated as quantum.
The parameter $\chi$ is depicted in Fig.~\ref{figure_7} as a function of time for a particle with impact parameter zero for three different models (without radiation reaction, LL equation and quantum description of radiation). As is clear from Fig.~\ref{figure_7}, although the naive model based on the NL equation predicts that $\chi\gtrsim 1$, but because of radiation reaction particle strongly decelerates on its way towards the focus and therefore $\chi$ remains small. Hence, one can expect that for our values of parameters LL equation provides qualitatively acceptable results.

In Fig.~\ref{b1}, the scattering angle  $\theta$ is shown as a function of the impact parameter. The studies are performed for $N_p = 1024$ particles distributed initially uniformly at $x = -18\pi$ with the initial momenta directed along $y$ axis and $\gamma_0 = 10^3$. The impact parameter corresponds to the initial coordinate $-12\pi<y<12\pi$ of the particles. The parametric studies presented in this section  are performed using the eRKN4 and iRKNp ($p=4,6,8$) methods with a time step value $h=10^{-4}\omega_0^{-1}$. To estimate the relative accuracy in a scattering angle $\theta$ we employed the iRKN8 method. The relative numerical errors  $|\theta_{4,6}-\theta_8|$ in the scattering angle $\theta$ are shown in Fig.~\ref{b2}. The errors in the scattering angle have pronounced dependency on the impact parameter. Thus the accuracy of the method is sensitive to the local gradients of the field. The error in the Minkowski norm for all the implicit methods under consideration is not sensitive to the impact parameter (see Fig.~\ref{b3}), but error of explicit eRK4 and eRKN4 methods shows a pronounced dependence on the impact parameter. However, the eRKN4 method performs slightly better than eRK4 in preservation of the on-shell condition. 

\section{Summary and conclusions}
\label{conclusion}

In this paper we studied the numerical methods for an accurate integration of the Landau-Lifshitz equation. 
The Landau-Lifshitz equation is proven to be a robust and computationally cheap model to include the radiation reaction effects into a particle dynamics. The result of our study showed advantage of the Runge-Kutta-Nystr\"om methods applied to the Landau-Lifshitz equation over the Runge-Kutta methods of the same order. The Runge-Kutta-Nystr\"om methods are specially designed to solve the second order ODEs with fewer amount of arithmetical operations than it is required for the equivalent system of the first order ODEs. One important observation is that the Minkowski norm is not always conserved in the course of the numerical simulation for both the Landau-Lifshitz and Newton-Lorentz equations. 
This problem is particularly strong for explicit methods, but the implicit methods keep the Minkowski norm bounded. 

We derived the four, six and eight order Runge-Kutta-Nystr\"om methods based on the Gauss-Legendre collocation points. These methods demonstrate excellent numerical accuracy and conservation properties. At first glance their numerical complexity is higher since they involve iterations. But in fact, implicit methods outperform the explicit ones since they allow larger integration steps. Further speed up 
of implicit RKN can be gained if more sophisticated iteration process is applied. 

The accuracy of all the methods under consideration were thoroughly tested against two analytical solutions of the Landau-Lifshitz equation. The first benchmark case was the solution of the Landau-Lifshitz equation for a charged particle rotating in a constant magnetic field. Radiation losses in this case correspond to the well known synchrotron radiation. The second analytical solution considered  was particle motion in a plane wave. We focused on a circularly polarized plane wave because in this case the phase can be explicitly expressed in terms of the proper time. 

Finally, we presented results of successful application of the RKN methods to simulation of scattering of high-energy electrons on focused laser beams. Our simulation indicates that taking into account radiation reaction results in strong deviation of scattering patterns from those predicted by the ponderomotive scattering theory. We also carried out quasiclassical modelling of radiation reaction due to emission of hard photons. The obtained results are qualitatively similar to those discovered with the Landau-Lifshitz equation. The results of the scattering test reveal  strong correlation between  the numerical accuracy and the impact parameter of a scattering particle. The local loss of accuracy happens due to strong gradients of the external field. To restore accuracy an adaptive time step control can be applied. This problem has to be attacked in future studies. 

We believe that our results for a Landau-Lifshitz equation solver will be useful in routine calculations for  planning future laser-matter interaction experiments in the range from moderate to high laser intensities. 

\begin{acknowledgments}
This research was supported by the Sonderforschungsbereiche TransRegio-18 (Project B13), by the Cluster-of-Excellence Munich-Centre for Advanced Photonics (MAP), Arnold Sommerfeld Zentrum (ASZ), Russian Fund for Basic Research (grant 13-02-00372) and the President program for support of young Russian scientists and leading research schools (grant NSh-4829.2014.2). We are grateful to Prof. N.B. Narozhny and Prof. A. Scrinzi for valuable discussions.
\end{acknowledgments}

\appendix
\section{Higher order collocation RKN method}
\label{irkn6}
 
In case $s=3$ the Lagrange basis of polynomials~(\ref{lagra}) takes the form
\begin{align}
 L_1 = \frac{\xi-c_2}{c_1-c_2}\cdot\frac{\xi-c_3}{c_1-c_3},\quad  L_2 = \frac{\xi - c_1}{c_2-c_1}\cdot\frac{\xi-c_3}{c_2-c_3}, \quad L_3 = \frac{\xi-c_1}{c_3-c_1}\cdot\frac{\xi-c_2}{c_3-c_2}.
\end{align}
Evaluation of the coefficients $A_{ij}$, $B_{ij}$, $a_i$ and $b_i$ requires just elementary integration. 
  \begin{table*}[!ht]
   \caption{Gauss-Legendre RKN method of order six.}
   \label{glrk6}
\subfloat{
      \begin{tabular}[b]{c|lcr}
      $\frac{1}{2} - \frac{\sqrt{15}}{10}$  &$\frac{5}{36}  $   				        & $\frac{2}{9}-\frac{\sqrt{15}}{15}$ &$\frac{5}{36}-\frac{\sqrt{15}}{30}$\\
      $\frac{1}{2}$    			         &$\frac{5}{36}  + \frac{\sqrt{15}}{24}$     &$\frac{2}{9}$                                &$\frac{5}{36}-\frac{\sqrt{15}}{24}$\\
      $\frac{1}{2} + \frac{\sqrt{15}}{10}$&$\frac{5}{36}  + \frac{\sqrt{15}}{30}$     &$\frac{2}{9} + \frac{\sqrt{15}}{15}$ &$\frac{5}{36}$\\\hline
      %%%--------------------
       &$\frac{5}{18}$				       &$\frac{4}{9}$ 					&$\frac{5}{18}$
      \end{tabular}
      }
      %%%
      \subfloat{
      \begin{tabular}[b]{|lcr}
      $\frac{1}{120}  $   				        & $\frac{1}{12}-\frac{\sqrt{15}}{45}$           &$\frac{13}{12}-\frac{\sqrt{15}}{36}$\\
      $\frac{5}{96}  + \frac{\sqrt{15}}{72}$             &$\frac{1}{48}$                                             &$\frac{5}{96}-\frac{\sqrt{15}}{72}$\\
      $\frac{15}{36}  + \frac{\sqrt{13}}{120}$         &$\frac{{1}}{12}+\frac{\sqrt{15}}{45}$         &$\frac{1}{120}$\\\hline
      %%---------------------------
      $\frac{5}{36}+\frac{\sqrt{15}}{36}$            &$\frac{2}{9}$                                                &$\frac{5}{36} - \frac{\sqrt{15}}{36}$
      \end{tabular}
      }
      \end{table*}
The Python script for evaluation of the iRKN6 method can be obtained from authors upon request. The advantage of symbolic calculation is analytical form of the coefficients formulated in terms of the collocation points $c_1,\; c_2,\;c_3$. For example, $A_{11}$ and $B_{11}$ are given by 
\begin{align}
A_{11} = \frac{c_{1} \left[2 c_{1}^{2} - 3 c_{1} \left(c_{2} + c_{3}\right) + 6 c_{2} c_{3}\right]}{6 (c_1-c_2)(c_1-c_3)},\quad%\\
B_{11} = \frac{c_{1}^{2} \left[c_{1}^{2} - 2c_1(c_2+c_3)+6c_2c_3\right]}{12(c_1-c_2)(c_1-c_3)}.
\end{align}

In order to obtain the method of the order six, the coefficients must be evaluated at the Gaussian points $c_1=1/2-\sqrt{15}/10$, $c_2 = 1/2$, $c_3 = 1/2 +\sqrt{15}/10$. The resulting coefficients for iRKN6 method are summarized in Table~\ref{glrk6}.

\bibliography{rk}
\end{document}